# SECURING IEEE 802.11G WLAN USING OPENVPN AND ITS IMPACT ANALYSIS

Praveen Likhar, Ravi Shankar Yadav and Keshava Rao M

Centre for Artificial Intelligence and Robotics (CAIR)
Defence Research and Development Organisation (DRDO)
Bangalore-93, India
`{praveen.likhar,ravi.yadav,keshava}@cair.drdo.in`

## ABSTRACT

*Like most advances, wireless LAN poses both opportunities and risks. The evolution of wireless networking in recent years has raised many serious security issues. These security issues are of great concern for this technology as it is being subjected to numerous attacks. Because of the free-space radio transmission in wireless networks, eavesdropping becomes easy and consequently a security breach may result in unauthorized access, information theft, interference and service degradation. Virtual Private Networks (VPNs) have emerged as an important solution to security threats surrounding the use of public networks for private communications. While VPNs for wired line networks have matured in both research and commercial environments, the design and deployment of VPNs for WLAN is still an evolving field. This paper presents an approach to secure IEEE 802.11g WLAN using OpenVPN, a transport layer VPN solution and its impact on performance of IEEE 802.11g WLAN.*

## 1. INTRODUCTION

Among the various wireless technologies, Wireless LAN (WLAN) comes out as a popular local solution because of its features like mobility, easy to setup, low cost and handiness. WLAN offers wireless Internet and Intranet access to users in various restricted geographical places known as hotspots such as airports, hotels, Internet cafes and college campuses. IEEE 802.11 a, b and g are the well known and established standards for WLAN. The IEEE 802.11g WLAN technology is one of the fastest growing segment of the communications market today. It provides always-on network connectivity without, of course, requiring a network cable. Home or remote workers can set up networks without worrying about how to run wires through houses that never were designed to support network infrastructure. WLAN components plug into the existing infrastructure as simply as extending a phone line with a wireless phone. By removing the need to wire a network in the home, the cost of adoption and benefit of mobility within the home and the low cost of components make wireless networking a low-cost and efficient way to install a home network. But many users of WLAN technology are not aware or concerned about the security implications associated with wireless networks. On the other hand, wireless adoption within the corporate and medium-sized businesses has been severely inhibited by security concerns associated with sending sensitive corporate data over the air. Unlike its wired network counterpart, where the data remains in the cables, the wireless network uses open air as a medium. This broadcast nature of WLAN introduces a greater risk from intruders.

In particular, with the evolution of wireless networking in recent years has raised the serious security issues [1], [2]. These security issues are of great concern for this technology as it is being subjected to numerous attacks [3], [4], and [5]. The most common attacks on wireless LANs are unwanted or automatic connection to the wrong network, man-in-the-middle attack with a fake Access Point (AP), theft of information by illegal tapping of the network, intrusion from open air, scrambling of the WLAN and consumption of device batteries.

The Wired Equivalent Privacy (WEP) is a standard security mechanism for IEEE 802.11g WLAN. When it was introduced, it was considered as a secured algorithm. But later it was found that it can be cracked easily [3], [6], [7], and [8]. VPN technology has been used successfully to securely transmit data in wired networks especially when using Internet as the medium. This success of VPN in wired networks and the inherent security limitations of wireless networks have prompted developers and administrators to deploy it in case of wireless networks. A VPN works by creating a tunnel, on top of a protocol such as IP. In this paper we evaluated the impact of OpenVPN [9], transport layer VPN solution, on performance of IEEE 802.11g WLAN.

This work is extension of our previous work [10]. The paper is organized in eight sections. Following this introductory section we give a brief description of WLAN standards. In the third section we explain the WEP weakness and vulnerabilities. The fourth section gives overview of VPN technology and its need for WLAN. In the fifth section we explain OpenVPN its working and comparison with WEP. The sixth section describes experimental details. The seventh section presents the experimental results and their analysis. The eighth section concludes the paper.

## 2. WIRELESS LAN STANDARDS

The IEEE 802.11 is a set of standards for wireless local area network (WLAN) computer communications in the 2.4, 3.6 and 5 GHz frequency bands [11]. The 802.11a, b, and g standards are the most common for home wireless access points and large business wireless systems.

The 802.11a is faster than 802.11b with a data transfer rates up to 54Mbps. As compare to 802.11b it can support more simultaneous connections and suffers less interference as it operates in 5GHz frequency band. However, among the three standards 802.11a has shortest range.

The 802.11b works in 2.4 GHz frequency band and support maximum transfer rate of 11Mbps. As compare to 802.11a it uses less expensive hardware and better in penetrating physical barriers. It is more susceptible to interference as its working frequency is used by many electronic appliances.

The 802.11g operates in 2.4GHz frequency band with maximum transfer rate of 54Mbps and have backward compatibility with 802.11b. Being operated in the 2.4GHz it also susceptible to interference. In practical scenario distance coverage by 802.11g is better than 802.11a but slightly less than 802.11b.

These WLAN standards are summarised in the Table 1.

Table 1.  IEEE 802.11 WLAN Standards.

| Parameter | 802.11a | 802.11b | 802.11g |
| --- | --- | --- | --- |
| Maximum operating speed | 54 Mbps | 11Mbps | 54Mbps |
| Working frequency band | 5 GHz | 2.4 GHz | 2.4 GHz |
| Modulation technique | OFDM | DSSS | OFDM |

| Maximum indoor distance coverage | 18 mts | 30 mts | 30 mts |
|---|---|---|---|
| Maximum outdoor distance coverage | 30 mts | 120 mts | 120 mts |
| Remarks | No interference ; less distance due to high frequencies | Interference from RF sources like cordless phones | Interference, backwards compatible with 802.11b |

## 3. WIRED EQUIVALENT PRIVACY (WEP)

The WEP is a privacy protocol specified in IEEE 802.11 to protect the link data transmitted in WLAN. It refers to the intent to provide a privacy service to wireless LAN users similar to that provided by the physical security inherent in a wired LAN. The WEP encryption uses the RC4 symmetric stream cipher with 40-bit and 104-bit encryption keys. Although 104-bit encryption keys are not specified in the 802.11 standard, many wireless AP vendors support them.

### 3.1. Security Issues with WEP

Security researchers have discovered potential attacks that let malicious users compromise the security of WLAN that use WEP [5], [7]. The following is a list of such attacks:

- Passive attacks to decrypt traffic, based on statistical analysis.
- Active attacks to inject new traffic from unauthorized mobile stations, based on known plaintext.
- Active attacks to decrypt traffic, based on tricking the access point.
- Dictionary-building attacks, after analyzing enough traffic on a busy network.

WEP has been widely criticized for a number of weaknesses [6], [8]:

- WEP is vulnerable because of relatively short IVs and keys.
- Authentication messages can be easily forged.
- IV Reuse Problem: Stream ciphers are vulnerable to analysis when the keystream is reused.
- Integrity Check value Insecurity: WEP uses a CRC for the integrity check. Although the value of the integrity check is encrypted by the RC4 keystream, CRCs are not cryptographically secure. Use of a weak integrity check does not prevent determined attackers from transparently modifying frames.
- Key Management: The WEP standard does not define any key-management protocol and presumes that secret keys are distributed to the wireless nodes by an external key-management service.

### 3.2. Tools available for attacking WLAN

The various popular tools for attacking the WLAN are listed in Table 2.

Table 2. Tools for Attacking WLAN.

| Tool | Operating System | Description |
|---|---|---|
| Aircrack [13] | Linux / Windows | It is a WEP key cracking program for use on 802.11 networks. The primary purpose for the program is to recover the unknown WEP key once enough data is captured. |
| Airpwn [14] | Linux | It is a tool for generic packet injection on an 802.11 network. |
| Airsnarf [15] | Linux | It is a simple rogue wireless access point setup utility to steal usernames and passwords from public Wi-Fi hotspots. |
| BSD-Airtools [16] | Linux | It contains a bsd-based wep cracking application, called dweputils. It also contains a AP detection application similar to netstumbler (dstumbler) that can be used to detect wireless access points and connected nodes, view signal to noise graphs *etc*. |
| Dsniff [17] | Linux | It is counterpart of NetStumbler |
| Dstumbler [18] | FreeBSD | It is counterpart of NetStumbler |
| Fake AP [19] | Linux | It generates thousands of counterfeit WLAN access points. |
| KisMAC [20] | MacOS | It is a free stumbler application for MacOS X. It puts network card into the monitor mode, completely invisible and send no probe requests. |
| Kismet [21] | Linux | It passively monitors wireless traffic and dissects frames to identify SSIDs, MAC addresses, channels and connection speeds. |
| MacIdChanger [22] | Windows | It is a MAC address spoofing tool. This is generally used to conceal the unique MAC id that is on every network adapter. |
| MacStumbler [23] | MacOS | It is a utility to display information about nearby 802.11b and 802.11g wireless access points. |
| Netstumbler [24] | Windows | It is a wireless access point identifier running on Windows. |
| Wep0ff [25] | Linux / Windows | It is a tool to crack WEP-key without access to AP by mount fake access point attack against WEP-based wireless clients. |
| WEPCrack [26] | Linux | It is a tool that cracks 802.11 WEP encryption keys by exploiting the weaknesses of RC4 key scheduling. |

| WEPWedgie [27] | Linux | It is a toolkit for determining 802.11 WEP keystreams and injecting traffic with known keystreams. The toolkit also includes logic for firewall rule mapping, ping scanning, and port scanning via the injection channel and a cellular modem. |
|---|---|---|
| Wifitap [28] | Linux | It allows users to connect to wifi networks using traffic injection. |

## 4. VIRTUAL PRIVATE NETWORK (VPN)

VPN technology provides the means to securely transmit data between two network devices over an insecure data transport medium. VPN technology has been used successfully in wired networks especially when using Internet as the medium. This success of VPN in wired networks and the inherent security limitations of wireless networks have prompted developers and administrators to deploy it in case of wireless networks. A VPN works by creating a tunnel, on top of a protocol such as IP. VPN technology provides three levels of security:

- **Confidentiality**: To provide the security against the loss of confidentiality, VPN provides a secure tunnel on top of inherently un-secure medium like the Internet. The data is encrypted before passing through the tunnel which provides another level of data confidentiality. If an attacker manages to get into the tunnel and intercepts the data, that attacker will only get encrypted data.

- **Integrity**: VPN uses integrity check mechanism such as hashing, message authentication code or digital signature to protect against the modification of data. It guarantees that all traffic is from authenticated devices thus implying data integrity.

- **Origin Authentication**: VPN provides mechanism for origin authentication by using cryptographic mechanism such as message authentication code or digital signature.

- **Replay Protection**: VPN also provides security against replay attack by using sliding window mechanism.

### 4.1. Need for VPN in Wireless Networks

The WLAN did not focus on security as a primary requirement. Generally the main focuses were on connectivity, throughput and other architectural and functional issues. As compared to wired networking the wireless networking is inherently more prone to attacks and less secure. Physical boundary for a wireless network cannot be confined. Although WEP is an existing security mechanism for WLAN, researchers have found many vulnerabilities in it. The WEP is also subjected to numerous attacks. These security issues of WLAN, lead the researchers, vendors and analysts to look for a solution to prevent these attacks. The tunnelling of data using VPN technology is a widely agreed robust protection against many threats and attacks.

## 5. OPENVPN

The OpenVPN is free and open source user space VPN solution which tunnels the traffic through transport layer using TCP or UDP protocol for encapsulation and transfer of data. It uses virtual network interface (VNI) for capturing incoming traffic before encryption and sending outgoing traffic after decryption. Security in OpenVPN is handled by the OpenSSL [12] cryptographic library which provides strong security over Secure Socket Layer (SSL) using standard algorithms such as Advanced Encryption Standard (AES), Blowfish, or Triple DES (3DES). The OpenVPN uses a mode called Cipher Block Chaining (CBC) which makes the cipher text of the current block dependent on the cipher text of the previous block. This prevents an attacker from seeing patterns between blocks with identical plaintext messages and manipulating one or more of these blocks.

The VNI appears as actual network interface to all applications and users. Packets of incoming traffic sent via a VNI are delivered to a user-space program attached to the VNI. A user-space program may also pass packets into a VNI. In this case the VNI injects these packets to the operating system network stack to sends it to the location mentioned in destination address field of the packets. The TUN and TAP are open source VNI. The TAP simulates an Ethernet device and it operates with layer 2 packets such as Ethernet frames. The TUN simulates a network layer device and it operates with layer 3 packets such as IP packets [29, 30].

In Figure 1, the working of OpenVPN is explained and in Figure 2, the data flow in OpenVPN environment is shown.

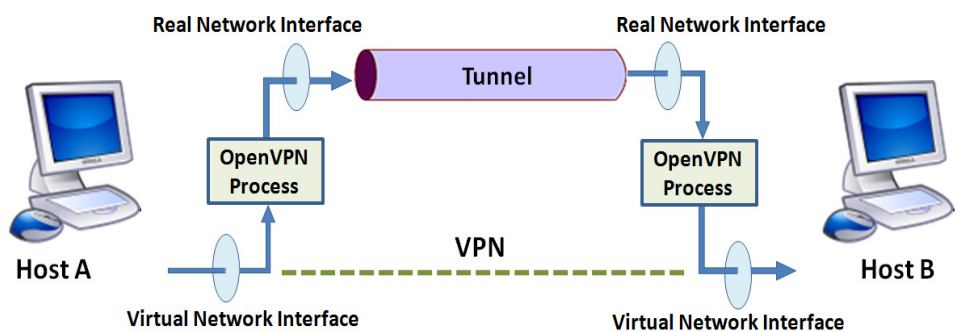

Figure 1. OpenVPN Tunnel between two end points

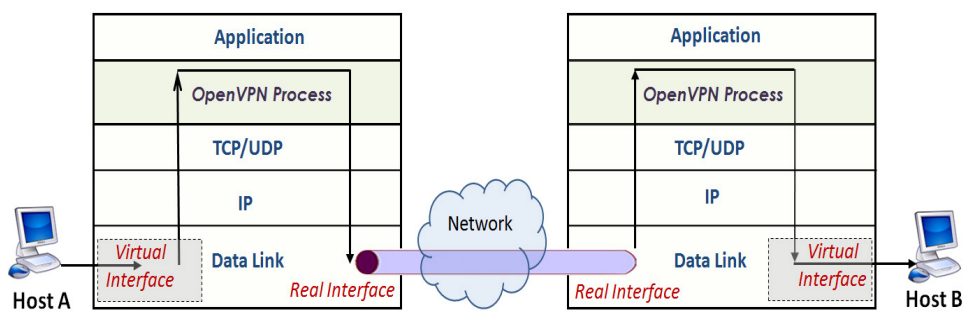

Figure 2. OpenVPN – Data Flow

The OpenVPN performs the following to secure the communications:

- Receives the packets of outgoing plain traffic from user space program by using the VNI.
- After receiving the packets, it compresses the received packets using Lempel-Ziv-Oberhumer (LZO) compression.
- After compression, it encrypts the packets using OpenSSL cryptographic library. For our experimentation we are using AES-128.
- OpenVPN also applies sliding window method to provide replay protection.

- Then it tunnels the packet using UDP or TCP protocol to the other end.
- On receiving the encrypted traffic at other end, the OpenVPN performs the reverse of cryptographic operations to verify integrity, authenticity etc.
- After successful completion of reverse cryptographic operations, it decompresses the packet.
- The decompressed packet is then passed via VNI to the user space program.

### 5.1. OpenVPN Cryptographic Operation

- OpenVPN uses a security model designed to protect against both passive and active attacks.
- OpenVPN security model is based on using SSL/TLS for session authentication and the IPSec ESP protocol for secure tunnel transport over UDP.
- OpenVPN uses the X509 PKI (public key infrastructure) for session authentication.
- OpenVPN uses TLS protocol for cryptographic key exchange.
- OpenVPN uses two factor authentication for authenticate the clients.
- OpenVPN uses OpenSSL cipher-independent EVP, an OpenSSL API that provides high-level interface to cryptographic functions, for encrypting tunnel data.
- OpenVPN uses HMAC-SHA1 algorithm for authenticating tunnel data.

### 5.2. Overcoming WEP vulnerabilities

In Table 3 shows how OpenVPN overcomes the WEP vulnerabilities by comparing it with WEP on various parameters.

Table 3. OpenVPN Comparison with WEP.

| Parameter | WEP | OpenVPN | Remark |
| --- | --- | --- | --- |
| Initialisation Vector (IV) | 24 bit (Too small) | Cipher –dependent and equal to cipher block size. | OpenVPN solves the IV reuse problem of WEP. |
| Encryption Algorithm | RC4 stream cipher | All Block Cipher supported by OpenSSL. Ex. AES, Blowfish, DES etc. | Encryption is fast and more secure in OpenVPN. |
| CBC Mode | Not supported | Supported | OpenVPN protects against know plain text attack. |
| Authentication | Open system and shared secret authentication | TLS based two factor authentication | OpenVPN authentication is strong than WEP. |
| Data Authentication and Integrity check | By using Cyclic Redundancy Check (CRC) | All OpenSSL authentication mechanism like HMAC-SHA1, MD5 etc. | OpenVPN provides better data authentication and integrity check. |

| Key Management | No key management | PKI X509 and pre shared secret | OpenVPN supports two established key management. |
|---|---|---|---|
| Replay Protection | No | Yes | OpenVPN uses sliding window mechanism to provide replay protection. |
| Attacks: Bit flipping, dictionary-building, FMS, etc. | Vulnerable to these attacks [5], [7], [8] | Secure against these attacks | OpenVPN provide security against well known WEP attacks. |

## 6. EXPERIMENT SETUP FOR PERFORMANCE MEASUREMENT

For analyzing the impact of OpenVPN on performance of IEEE 802.11g WLAN we created two experiment scenarios. First was for measuring the throughput under normal conditions and the second was to analyze the variation of traffic throughputs over an IEEE 802.11g WLAN when OpenVPN is implemented in WLAN. The following parameters were used as metrics for performance measurement during our experiments:

- **Throughput** is the rate at which bulk of data transfers can be transmitted from one host to another over a sufficiently long period of time.
- **Latency** is the total time required for a packet to travel from one host to another, generally from a transmitter through a network to a receiver.
- **Frame loss** is measured as the frames transmitted but not received at the destination compared to the total number or frames transmitted.
- **IP Packet delay variation** is measured for packets belonging to the same packet stream and shows the difference in the one-way delay that packets experience in the network.

### 6.1. Standard followed for performance measurement

We followed the IP Performance Metrics (IPPM) RFC 4148 [31], to measure the performance. The following is the list of metrics we have used along with the standard followed to measure these metrics.

- Maximum throughput achieved as per RFC 2544 [32],
- One-way Delay as per RFC 2679 [33],
- One-way Packet Loss as per RFC 2680 [34],
- IP Packet Delay Variation Metric as per RFC 3393 [35].

### 6.2. Requirements for Experimentation

The following is a list of the general Software and Hardware requirements for our experiments:

- Two laptops loaded with Red Hat Enterprise Linux 5,
- Ethernet Cables,
- TL-WA601G 108M TP-Link Wireless Access point,
- SPT-2000A Spirent test center.

### 6.3. Experiment Setup

In our experiment setup two laptops are connected using TP-Link Access point. The distance between the AP and the laptops is set to 4 meters to keep the signal strength high. Port-1 of Spirent test center is connected to laptop-1 and port-2 of Spirent is connected to laptop-2 using Ethernet cables of length 3 meters. These ports act as clients for laptops. These ports are used for traffic generation and analysis purpose. Port-1 of Spirent test center is used to generate the desired traffic for various data rates, frame sizes etc. Port-2 receives the traffic and analyses it. The analysis includes max throughput achieved, latency and packet delay variation with respect to various frame sizes. The 802.11g WLAN standard does not have inbuilt compression feature. OpenVPN supports both modes without compression and with compression, in our study we experimented both modes.

#### 6.3.1. Performance without OpenVPN

The experiment setup for this is shown in Figure 3. We carried out this experiment for measuring the baseline performance of IEEE 802.11g WLAN.

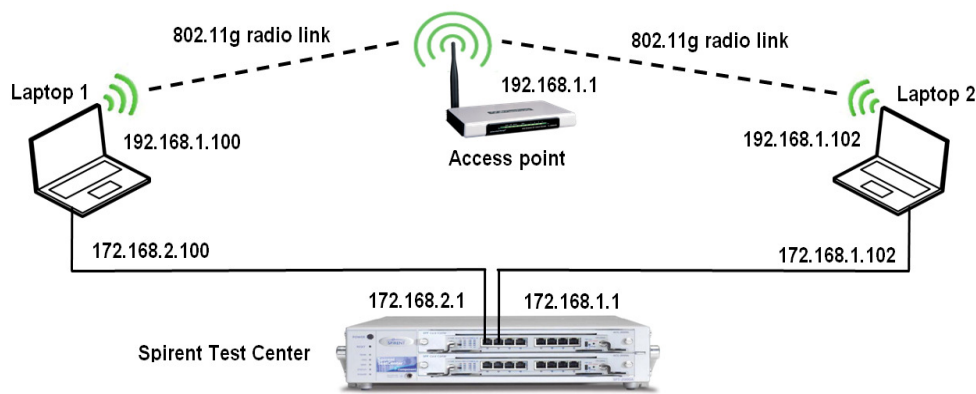

Figure 3. Experiment setup without OpenVPN

This experiment comprises of two steps. The first step measures the throughput with respect to UDP traffic, while the second step measures the throughput with respect to TCP traffic. In the first step, port-1 of Spirent test center sends UDP traffic of different frame sizes to laptop-1, which is connected to laptop-2 through wireless link using an Access Point (AP). Laptop-1 forwards this data to laptop-2 through AP and then laptop-2 send this data to port-2 of the Spirent test center. The second step of the experiment was conducted using the same environment variables described above, but this time TCP traffic was generated using port-1 to send traffic with different frame sizes from laptop1 to laptop2. We varied the size of the frame from 512 bytes to 1518 bytes.

#### 6.3.2. Performance with OpenVPN

Now our next aim is to analyze the impact of applying OpenVPN security solution to 802.11g WLAN. In this scenario first we have to run our OpenVPN solution on both the laptops. OpenVPN configuration files [36] for both laptops are given below in Table 4.

Table 4. OpenVPN configuration files.

| Configuration file – Laptop-1 | Configuration file – Laptop-2 |
|---|---|
| Port 5002 | Port 5002 |
| Proto udp | Proto udp |
| Dev tun0 | Dev tun0 |
| Remote 192.168.1.102 | Remote 192.168.1.100 |
| Ifconfig 20.20.20.1 20.20.20.2 | Ifconfig 20.20.20.2 20.20.20.1 |
| Cipher AES-128-CBC | Cipher AES-128-CBC |
| Secret static.key | Secret static.key |
| Comp-lzo | Comp-lzo |
| Keepalive 5  20 | Keepalive 5  20 |
| Persist-tun | Persist-tun |

The setup for this experiment is shown in Figure 4. To analyze the impact of applying OpenVPN security to 802.11g WLAN on the throughput of UDP and TCP traffic in IEEE 802.11g WLAN, we performed the experiments in two steps. In first step we measured the impact on UDP traffic over IEEE 802.11g and in second step we measured the impact on TCP traffic over IEEE 802.11g. Experimentation was carried out in the same manner as for baseline performance measurement.

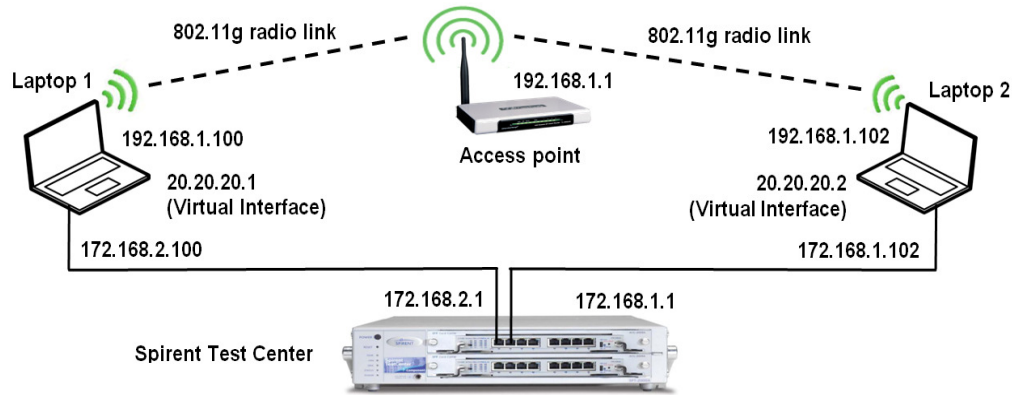

Figure 4. Experiment setup with OpenVPN

## 7. EXPERIMENT RESULT AND ANALYSIS

The results for all test scenarios of our experiment were collected from the test bed illustrated in the experiment setup section. Each experiment was repeated for twenty iterations to find the average performance values.

### 7.1. Throughput

The UDP and TCP throughput are measured as per RFC 2544 standards for different frame sizes. The results of these experiments for UDP are presented in Table 5 and in Figure 5. The results of these experiments for TCP are presented in Table 6 and in Figure 6.

Table 5. UDP throughput results.

| Frame Size (bytes) | UDP Average Throughput (Mbps) | | |
|---|---|---|---|
| | Without OpenVPN | With OpenVPN | |
| | | Without compression | With compression |
| 512 | 3.847 | 3.627 | 5.429 |
| 1024 | 5.429 | 4.574 | 11.238 |
| 1280 | 6.062 | 5.389 | 13.915 |
| 1518 | 6.906 | 6.062 | 16.09 |

Table 6. TCP throughput results.

| Frame Size (bytes) | TCP Average Throughput (Mbps) | | |
|---|---|---|---|
| | Without OpenVPN | With OpenVPN | |
| | | Without compression | With compression |
| 512 | 3.135 | 2.601 | 4.796 |
| 1024 | 4.796 | 4.065 | 10.929 |
| 1280 | 5.429 | 4.961 | 12.936 |
| 1518 | 6.062 | 5.62 | 16.09 |

Figure 5 and Figure 6 indicates that the throughput increases for both UDP and TCP traffic with increased frame size. Throughput increases because when the data transmitted using large frames the total overhead for transmitting the data due to frame headers will be less as compare to when the data is transmitted using small frames. The throughput is decreased slightly when OpenVPN is applied because of the increased overhead which is due to encapsulation and cryptographic operations used by OpenVPN. When compression is used with OpenVPN throughput increases since compression reduces the packet size in physical interface. Throughput in this case is better than the throughput in normal case i.e. without OpenVPN because IEEE 802.11g does not has inbuilt compression and after compression packet size reduces considerably if data is not randomly distributed which is true most of the time.

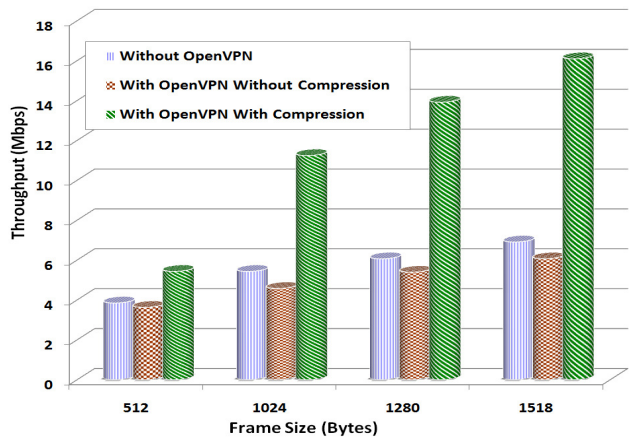

Figure 5. UDP throughput according to frame size.

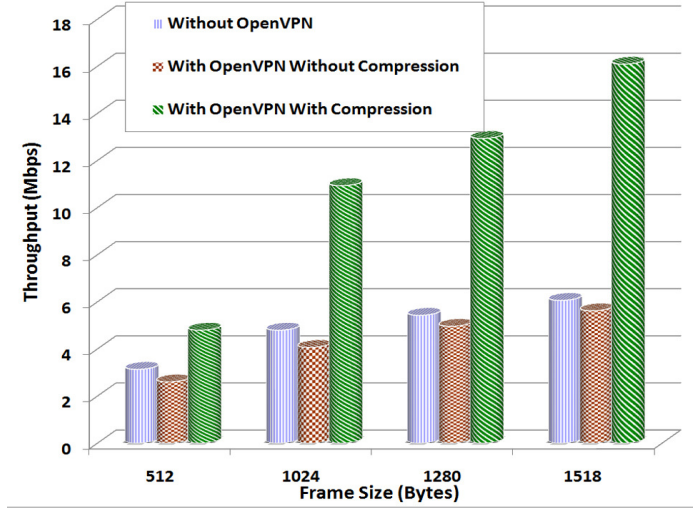

Figure 6. TCP throughput according to frame size.

As mentioned above, throughput of WLAN decreases slightly for both UDP and TCP traffic after applying OpenVPN security. Table 7 lists the decrease in throughput corresponding to each frame size and for UDP and TCP traffic. From the table it is clear that maximum decrease in throughput is 15.84% (0.86 Mbps) in case of UDP traffic. In case of TCP traffic the maximum decrease in throughput is 16.91% (0.53Mbps).

Table 7. Loss in throughput after applying OpenVPN without compression

| Frame Size (bytes) | UDP | | TCP | |
|---|---|---|---|---|
| | Loss (Mbps) | Loss % | Loss (Mbps) | Loss % |
| 512 | 0.22 | 5.72 | 0.53 | 16.91 |
| 1024 | 0.86 | 15.84 | 0.73 | 15.22 |
| 1280 | 0.67 | 11.05 | 0.47 | 8.66 |
| 1518 | 0.84 | 12.16 | 0.44 | 7.26 |

## 7.2. Average latency

We measured the average latency as per RFC 2679 standards for both UDP and TCP traffic with various frame size. Figure 7 and Figure 8 shows the experimental result for average latency. The figures clearly indicate that the latency increases for both UDP and TCP traffic as we increase the frame size. This is due to the fact that round trip time is proportional to the size of frame. From these figures it is also clear that latency is less for normal case as compare to the two cases of OpenVPN mode because in OpenVPN mode additional processing is required for performing cryptographic operation, compression and encapsulation. From the experiment results we also analyzed that the latency in case of OpenVPN without compression is more than in case of OpenVPN with compression. Even though compression takes some processing time it reduces the frame size which results in decreased transmission time as compare to the transmission time when frame is not compressed.

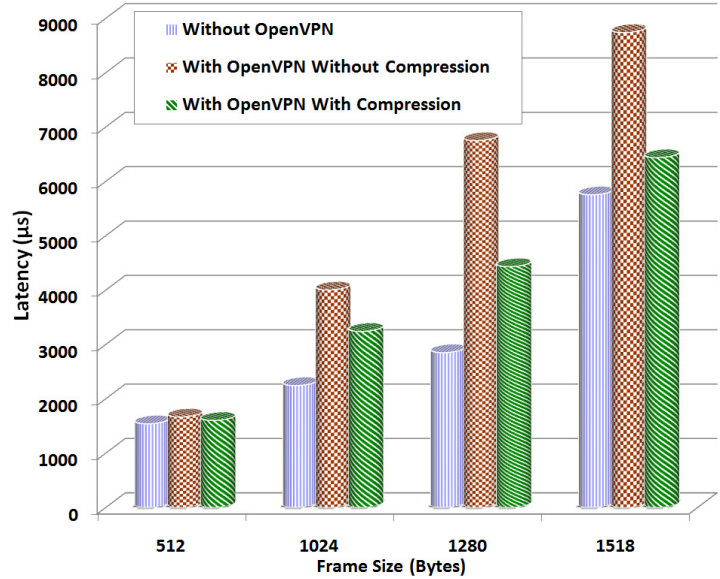

Figure 7. UDP Average Latency according to frame size.

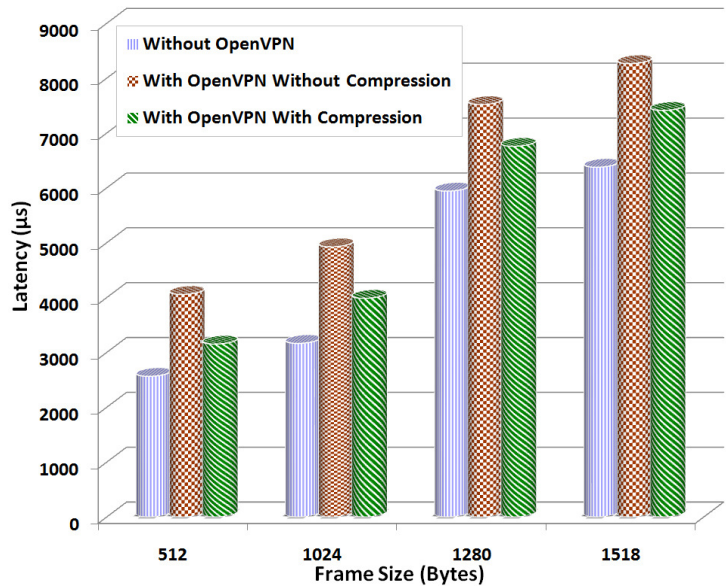

Figure 8. TCP Average Latency according to frame size.

### 7.3. Frame Loss Percentage

Frame Loss Percentage is measured as per RFC 2680 standards for UDP and TCP traffic with different loads. The results of these experiments are presented in Figure 9 and Figure 10. These figures indicate that as we increase the load the Frame Loss Percentage increases for both UDP and TCP traffic. The frame loss percentage increases exponentially as the load crosses the throughput value corresponding to particular frame size.

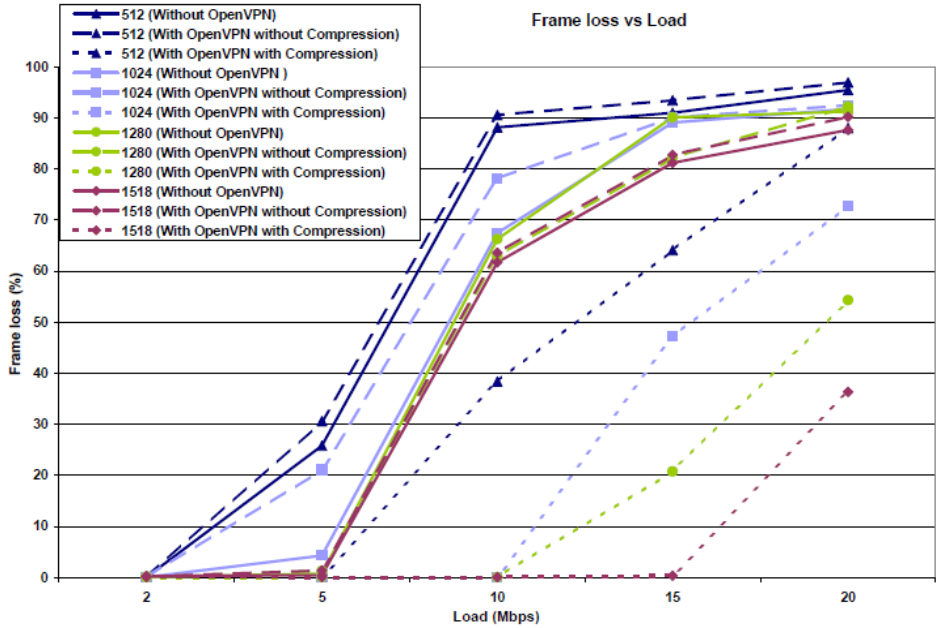

Figure 9. UDP Frame Loss Percentage according to frame size.

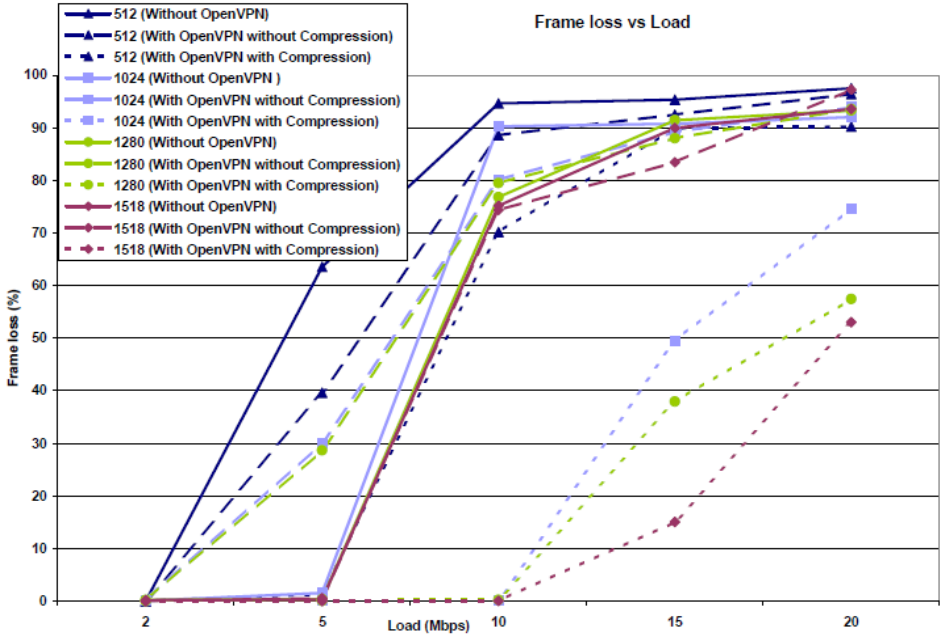

Figure 10. TCP Frame Loss Percentage according to frame size.

### 7.4. IP Packet Delay Variation

The IP Packet delay variation is measured as per RFC 3393 standards for UDP traffic for different frame size with different transmission rates. The result of this experiment is presented in Figure 11. This figure indicates that as we increase the load the IP Packet delay variation increases. From the above figures we observe that with the use of compression with OpenVPN,

110

International Journal of Network Security & Its Applications (IJNSA), Vol.3, No.6, November 2011

the IP Packet delay variation decreased as compared to normal case because compression reduces the payload size of packet.

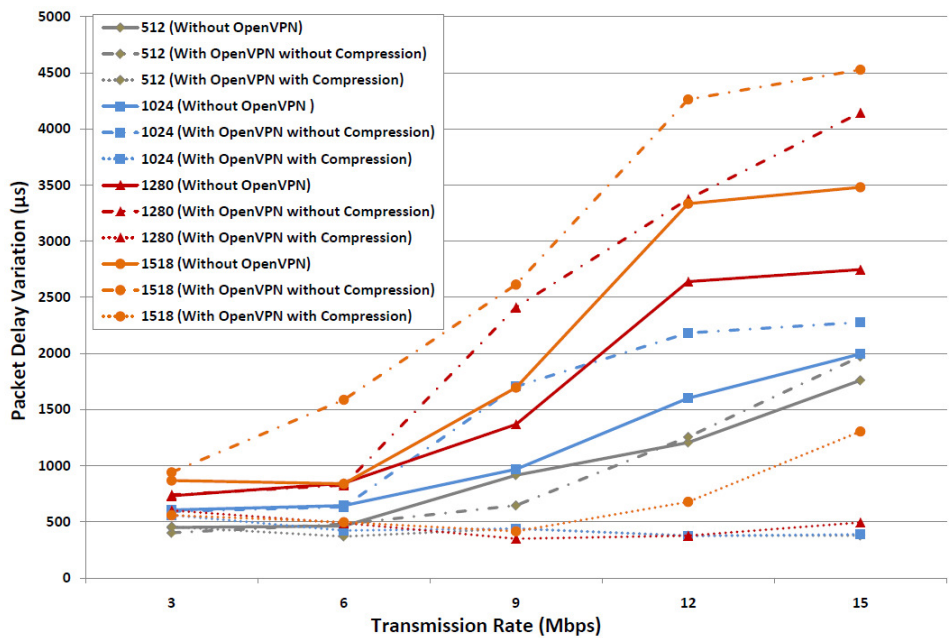

Figure 11.  TCP Frame Loss Percentage according to frame size.

## 8. CONCLUSIONS

The benefit of wireless networks is driving the explosive growth of the WLAN market, where security has been the single largest concern for wireless network deployment. Through this paper we discuss why security is a major concern for WLAN. We have investigated and listed security vulnerabilities and attacks on the standard security mechanism for WLAN called WEP. We also explained how VPN can be used as a security solution for WLAN. In this work a transport layer tunneling based VPN solution named OpenVPN was adopted and implemented for 802.11g WLAN. To show how OpenVPN overcomes the weaknesses of WEP, we have compared OpenVPN with WEP based on various security parameters. The performance analysis was carried out with respect to throughput, latency, frame loss and IP packet delay variation. To measure these performance matrices we have followed RFC4148, RFC 2544, RFC 2679, RFC 2680 and RFC 3393. Experimentation was carried out for both UDP and TCP traffic with respect to various data rates and frame sizes using Spirent test center to analyse the impact of OpenVPN on performance of 802.11g WLAN. From the experimental results we can conclude that there is slight decrease in performance of 802.11g WLAN with the implementation of OpenVPN. But there is an increase in the performance of 802.11g WLAN with the use of compression in OpenVPN.

## ACKNOWLEDGEMENTS

We would like to thank Director CAIR for supporting us to work in this area. We would also like to thank Dr. G. Athithan for his help and constructive suggestions throughout.

International Journal of Network Security & Its Applications (IJNSA), Vol.3, No.6, November 2011

**Authors**

**Praveen Likhar**, obtained bachelor's degree from Maulana Azad National Institute of Technology, Bhopal. He is working as a scientist with Centre for Artificial Intelligence and Robotics (CAIR), Defence Research and Development Organisation (DRDO), Bangalore. His research spans computer network security, communication security and wireless LAN security.

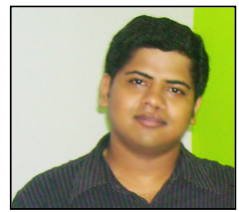

**Ravi Shankar Yadav**, obtained master's degree from Indian Institute of Science (IISc), Bangalore and his bachelor's degree from Motilal Nehru National Institute of Technology, Allahabad. He received "BEL R&D Excellence Award" in the year 2006 and also awarded with "Lab Group Technology Award" in the year 2007. He is working as a scientist with Centre for Artificial Intelligence and Robotics (CAIR), Defence Research and Development Organisation (DRDO), Bangalore. His research interests are cyber security, information security and computer networks.

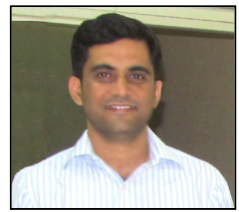

**Keshava Rao M,** obtained Master's degree from BITS, Pilani. He received BEL R&D excellence award in the year 2006 and currently he is working as a scientist with Centre for Artificial Intelligence and Robotics (CAIR), Defence Research and Development Organisation (DRDO), Bangalore. He is having more than 10 year working experience in the area of information security.

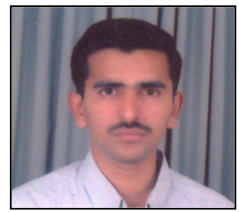